# On topological aspects of 2D graphene like materials

M. J. I. Khan[*¶], Kamran Tahir[*], S. Babar[*]

Laboratory of theoretical and computational physics

[*]Department of Physics, Bahauddin Zakariya University, 60800, Multan, Pakistan.

[¶]International Islamic Youth Academy, Multan, Pakistan.

Email: drjunaid.iqbalkhan@bzu.edu.pk

mjik.pph@gmail.com

## Abstract

We study the graphene lattice with a curvature effect. The action depicting multilayers of graphene is portrayed in curved spacetime and effective Dirac equation scopes the curvature effect. The magnetic field is responsible for the geometric variations and these changes are identified as topological aspects in graphene. By varying the geometry of the graphene one can create topologically distinct surfaces which could be remarked as torus or sphere. The 2D hexagonal tessellation induces a curvature effect and the tessellation plane is reduced to a 2-torus having genus $g = 1$.

# Contents

## 1. Introduction

Graphene is a system of carbon atoms forming a hexagonal lattice delibrating a system in which carbon atoms are $sp^2$ hybridized. Since it experimental realizations in 2004, most of the peoples have studied the subject of graphene with and without magnetic field [1, 2]. In an empirical realization this material is simply a monolayer of carbon crystal forming a two dimensional hexagonal lattice with a geometric structure [3]. The experimental interest in graphene is a matter of interest because it replaces the silicon based technology and provide a space for technology development in term of nanotechnology. Graphene get theoretical mileage beacause dynamics of fermions are described by Dirac equation in two dimensions and charge careers in graphene mimic relativistic particles with zero rest mass and have an effective speed comparable to speed of light [4]. In this way, low energy excitations are described by a four component dirac fermion [5]. The two dimensional system of hexagonal lattice is fascinating and subtle in the sense because it go through the knowledge of Quantum electrodynamics. The core issues of technology provide a bridge between condensed matter and high energy physics [6-8] which allows a better understanding of band structure and electronic properties. The special behaviour of graphene originates on topology and geometry of the honeycomb lattice and have a profound implications on transport and optical properties.

To study the topology of 2D materials like graphene [9], we have to study curvature effects in the lattice. When talking about graphene, low energy description of curved surface can be described by Dirac equation that reflects the fermionic behavior inside the lattice of 2D materials [10, 11]. The most common example of 2D materials could be graphene which is most common in both experimental and theoretical scope [12, 13]. Theoretical study of graphene has shown tremendous results under the magnetic field that reveals both fractional and integral quantum hall effects in graphene.

In this paper, we discuss the effective Dirac equation in case of the curvature effects. This paper is organized as follows. In section 2, we mention an action depicting a multilayer graphene system and encoding informations about gauge field and Dirac equation. We state dirac equation for curved spacetime. Section 3 interprets the effective Hamiltonian of graphene which is portrayed by dirac equation in curved spacetime and involving a remark about the non-abelian gauge field. Topology is addressed in section 4, where the variation in geometry in graphene lattice is studied which results a torus after a specific geometrical change.

## 2. Curved Spacetime

Let's consider a flat lattice of graphene which is defining a 2D plane where $k$ and $\acute{k}$ are the $k$ vectors. Physics of graphene is natural when graphene lattice is simply defined and can be modeled by a Dirac equation. The Dirac like description of the energy band diagram of graphene is topological and for a given $k$ there is only one spin orientation. The topological behavior of graphene can exhibit many interesting facts like edge states, geometry of lattice [14]. Variation of geometry can create topologically distinct configurations which can be identified topologically [15]. The topology in graphene lattice can be introduced by the effect of an external field which could be magnetic or electric field. Under the effect of external magnetic field, the curvature is produced in graphene lattice and physically this curvature induces an effective field that is to be mentioned in Dirac equation. The field labeled as the gauge field that is a consequence of curvature. The low energy dynamics of the system can be described by the action,

$$J = \int \frac{1}{4g^2} F_{\mu\nu}^2 + \frac{1}{4g^2\xi}(\partial_\mu A^\mu)^2 + \sum \bar{\psi}^a D_\mu \psi^a \qquad (1)$$

The above action best describes the electrodynamics of graphene at weak coupling. This action depicting the notion of fermions localized in (2+1) dimensions and it described by Dirac equation. The second term carries gauge field $A^\mu$ which is containing a gauge group U(1) associated with the lattice. It is a covariant gauge fixing term. Of course, $F_{\mu\nu}$ is interpreted as the legtitimate field strength related to unbroken $U(1)$ part of the gauge symmetry.

The third term is a Dirac like equation which is defined for multi layers of graphene.

$$\sum \bar{\psi}^a \gamma^\mu (i\partial - A)\psi^a \qquad (2)$$

Where $\psi^a$ is a complex $2 \times 2$ complex two dimensional Spinors with $a = 1, 2, \ldots\ldots, k$. Where $\gamma$ matrices are given in term of $2 \times 2$ matrices where $\gamma^\mu$ matrices satisfying Clifford algebra in (2+1) dimensions [16]. We consider significant terms,

$$\frac{1}{4g^2\xi}(\partial_\mu A^\mu)^2 + \sum \bar{\psi}^a \gamma^\mu D_\mu \psi^a \qquad (3)$$

The physical significance of gauged Dirac operator $D = \gamma^\mu(i\partial - A) \equiv \gamma^\mu D_\mu$ is that it describes theory with fermions $\psi^a$ in the background of external field with a definite topological charge $Q_{top}$ [17]. The gauge field is significant in the sense that it gives the impact on the curvature that define physical properties of graphene. The connections between gauge field topology and axial anomaly has been discussed in [18]. Three dimensional topological field theories are nicely related to two dimensional physics and they give geometrical realizations have been addressed in [19]. The two dimensional topological theories are mostly described by Chern Simons (CS) theory that encodes the information's about the gauge field. In [20] M. J. I. Khan et al. have established the theory of Quantum Hall effect by developing the CS terms in (2+1) dimensions. The system restores on gauge field that is associated with the lattice without external field and the other gauge field is produced due to the distortion. This gauge field is labeled as the effective gauge field [21]. The magnetic field $\boldsymbol{B} = \boldsymbol{\nabla} \times \boldsymbol{A}$ could produce positive curvature. The magnetic field along the boundary can be written as,

$$\int B = \oint \boldsymbol{\nabla} \times \boldsymbol{A} = 2\pi \qquad (4)$$

This magnetic field produce distortion in the lattice which appears to be the identification of different topological configurations. The effect of gauge field $\nabla \times \mathbf{A}$ is dominant over the boundry and it gives contributions to the magnetic charge. So integration over the surface without boundary becomes meaningless [21]. It is supported by the fact that for topological insulators the conduction is present at the edge states.

The effective action for the 2D system describing the curvature effect is written in term of zweibein. We can reconstruct the metric for flat space from zweibein $g_{\alpha\beta} = \delta_{ab} e_\alpha^a e_\beta^b$. The physical importance of the zweibein is that they transform Pauli matrices from flat to curved spacetime. By the introduction of zweibein, we insert an additional $U(1)$ symmetry that is related to the effective gauge field $A$ where the $U(1)$ is a compact group. The $\gamma^\mu$ matrices in (2+1) curved spacetime can be written as,

$$\gamma^\mu = \gamma^a e_a^\mu \tag{5}$$

The supersymmetry functional of the theory is $(\phi, \psi, e, \chi)$ where $\chi$ is a supersymmetric partner of zweibein called gravitino. The wave function transforms the spinor as,

$$\psi \mapsto e^{\varsigma^{\mu\nu} \sigma_{\mu\nu}} \tag{6}$$

Where $\varsigma^{\mu\nu}$ is a function of spacetime. The above action in curved spacetime can be written as,

$$J = \int \frac{1}{4g^2} F_{\mu\nu}^2 + \frac{1}{4g^2 \xi}(\partial_\mu A^\mu)^2 + \sum i \gamma^\mu e_a^\mu D_\mu \psi^a - m\psi^a \tag{7}$$

The above equation is a coupling of matter field to gauge field where $e_a^\mu$ zweibein field and the gravitational spin connection. The zweibein defines a local rest frame, allowing the constant Dirac matrices to act at each spacetime point.

### 3. Curvature Effect

There are numerous ways to produce curvature effect [21-24] in graphene tessellation. It can be observed by having a dissection of a unit cell. From physical point of view, the curvature effect plays a magic in defining the theoretical framework. It induces an effective gauge field which is associated with the gauge group $U(1)$. The non-abelian gauge field depicts the effective Hamiltonian description of curved graphene. Abelian gauge field satisfies [22],

$$\oint_c A \cdot dr = \pi/2 \tag{8}$$

Where $\mathbf{B} = \nabla \times \mathbf{A}$ is magnetic flux. The effective Hamiltonian of the curved graphene will be,

$$H = \begin{pmatrix} -ie_a^\mu \sigma^k (\nabla_\mu - ieA_\mu) & 0 \\ 0 & ie_a^\mu \sigma^k (\nabla_\mu + ieA_\mu) \end{pmatrix} \tag{9}$$

Where $\nabla_\mu$ is an operator which reduces to partial derivative in flat space with Cartesian coordinates but transform as a tensor on arbitrary manifold which in this case is curved graphene. It is called covariant derivative. It describes the curvature effect on spinors.

## 4. Topological Effects

Topology is interplay with the geometric variations in various surfaces. Topology is a branch of mathematics which is associated with the global properties of geometrical objects. By the continuous deformations, we can bring the topological change but these properties are unaffected with these deformations. Depending upon the dimensions, we can observe complicated topologies but the simple one topology can be studied in 2D where usually the surfaces are compact and can be a circle or torus. Physically topology can play a role to study gauge field associated with each node of the lattice. If topology is caused by the external fields then curvature effects are also observed on surface in 2D and gravitational effects that counter scalar fields. So topology helps us to study gauge field and scalar fields. Curvature plays a significant role in developing topological interplay. Geometric variation can produce topology and we can indentify different topological configurations such as sphere, circles or torus on which we can define effective dirac equation. Physically these surfaces are marked as nanotubes, buckyballs. Nanotube is produced from a torus by cutting it along its smallest circle. Curvature can be positive or negative. Negative curvature is observed at saddle points and can be introduced by heptagons. By interplay between geometric variations and topology we can study different forms of graphene sheet in one dimension less than 2D. Liu [25] and Martel et al. [26] have reported carbon nanotubes forming ring like structures suspected to be toroidal [27]. These toroidal structrues are topological and these nanotorus have theoretically shown to possess unusal electronic magnetic properties like serving as a propotype of quantum wire ring [28]. A very few theoretical studies are done about the structures containing nanotori that how the curvature effects its local structures [29]. It can be studied theoretically by small scale molecular dynamics where in small systems the curvature in the structure of graphene is induced by locally adding pentagons and heptagons into the structures which are likely to be hexagonal. Another way is to consider combine nanotubes of different helicity. But for large scale molecular dynamics it is addressed by the molecular simulation techniques, which we expect to give some insight to the structure of carbon nanotubes under uniform bending strain.

Here we are interested in studying the geometric variations on tessellation or hexagonal lattice which is simply a two dimensional plane. A hexagon is bordered by six other hexagons and further hexagons. A regular hexagon can be equivalent to a torus by certain geometric procedure. First we have to identify the edges of the hexagon that would help us to connect the opposite edges while stretching and shriking. The geometric changes that we will produce will be equivalent to the flat torus. We take the single hexagon and rotate it by 90° along z-axis. We roll up the opposite vortices along y-axis. It resembles like a tube and after twisting of this tube by 180°, we will have finally a hexagonal two-torus. It is a compact surface and is a simple example of simply connected space.

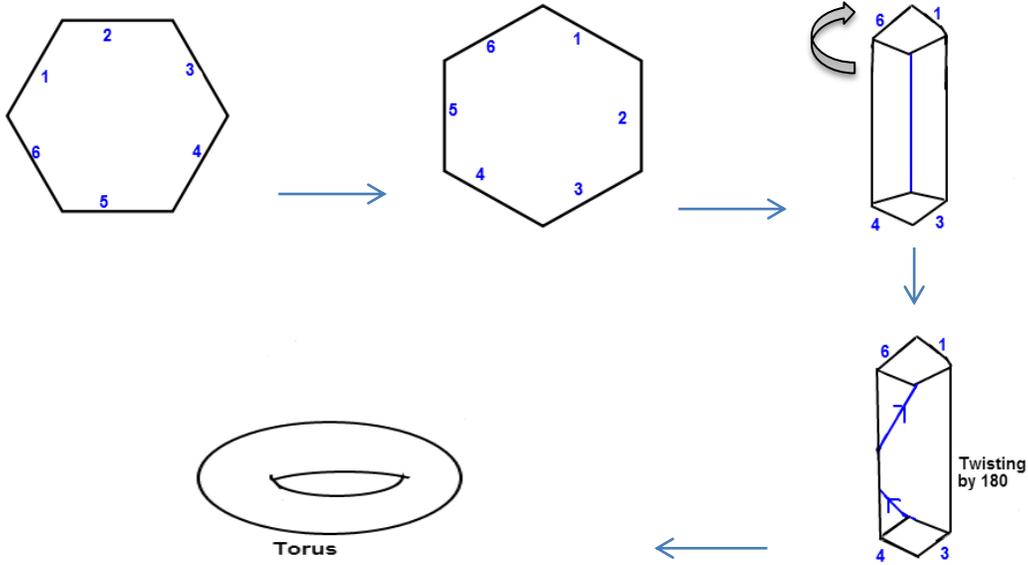

Fig. 1 (Geometric changes in hexagon)

Mathematically, torus is a product of n-circles.

$$T^n = s^1 \times s^1 ....\times s^1 \qquad (10)$$

But a two-torus is $T^2 = s^1 \times s^1$. It is a fact that topological surfaces are identified with a genus which is equivalent to number of holes. So torus is compact surface with $g = 1$.

## 5. Conclusion

The action depicting curvature effect is studied and it gives theoretical novelty of graphene multi-layers. Curvature effect in graphene is studied due to magnetic field. The action containing dirac equation is modified which include the gravitational terms like zweibein. The Hamiltonian of the curved graphene is a $2 \times 2$ matrix. Topological aspects are observed while geometrical changes in hexagon and these are identified as the torus. The topological discussion of gauge field is yet a question of interest because one need to introduce an additional gauge field due to effective curvature and curved spacetime.